\documentclass[12pt]{article}

\usepackage{graphics}
\usepackage{amssymb}
\usepackage[driver]{graphicx}

\newcommand{\QED}{\mbox{\rule[-1.5pt]{6pt}{10pt}}}
\newcommand{\C}{\mathbb{C}}
\newcommand{\N}{\mathbb{N}}
\newcommand{\R}{\mathbb{R}}
\newcommand{\Z}{\mathbb{Z}}
\newcommand{\LL}{{\cal L}}
\newcommand{\OO}{{\cal O}}

\newtheorem{theorem}{Theorem}[section]
\newtheorem{lemma}{Lemma}[section]
\newtheorem{remark}{Remark}[section]

\begin{document}

\title{Magnetic layers with periodic point perturbations}
\date{}
\author{P.~Exner,$^{1,2}$ K.~N\v{e}mcov\'{a}$^{1,3}$}
\maketitle
\begin{quote}
{\small \em 1 Nuclear Physics Institute, Academy of Sciences,
25068 \v Re\v z \\ \phantom{a)}near Prague, Czech Republic \\ 2 Doppler
Institute, Czech Technical University, B\v rehov{\'a} 7,
\\ \phantom{a)}11519 Prague, Czech Republic \\ 3 Institute of Theoretical Physics,
FMP, Charles University, \\ \phantom{a)}V Hole\v{s}ovi\v{c}k\'ach
2, 18000 Prague, Czech Republic
\\
\phantom{3)} \rm exner@ujf.cas.cz, nemcova@ujf.cas.cz}
\vspace{8mm}
\end{quote}

\begin{abstract}
\noindent We study spectral properties of a spinless quantum
particle confined to an infinite planar layer with hard walls,
which interacts with a periodic lattice of point perturbations and
a homogeneous magnetic field perpendicular to the layer. It is
supposed that the lattice cell contains a finite number of
impurities and the flux through the cell is rational. Using the
Landau-Zak transformation, we convert the problem into
investigation of the corresponding fiber operators which is
performed by means of Krein's formula. This yields an explicit
description of the spectral bands, which may be absolutely
continuous or degenerate, depending on the parameters of the
model.
\end{abstract}



\section{Introduction}

The object of the present study is a three-dimensional spinless
quantum particle interacting with a homogeneous magnetic field and
a periodic point potential. In addition, the particle is confined
to a flat layer of a constant width $d$ with Dirichlet boundary
conditions, the magnetic field being perpendicular to the layer.
The lattice of the point interactions is two-dimensional, not
necessarily planar, and both of its generating vectors are, of
course, parallel to the boundary planes of the layer.

The problem is clearly motivated by the need of modeling electrons
in a semiconductor layer with `impurities', either natural or
artificially created. If they may be supposed to be well
localized, one often describes them by point interactions. It is
useful because in such a way we get solvable models which make it
possible to derive the spectral and scattering properties of a
given configuration in a relatively simple way, and at the same
time they reproduce the basic features of an actual
crystal-lattice layer with some alien atoms and a low-density
electron gas. Various systems of this type have been investigated
in the literature; we refer to our previous paper \cite{EN} and an
earlier study of the two-dimensional analogue \cite{EGST} for an
extensive bibliography as well as for a discussion of the used
approximations.

The said paper \cite{EN} was devoted to investigation of a flat
hard-wall layer with a finite number of point interactions, with
or without a magnetic field; we analyzed there the spectra, and in
the former case also the scattering properties of such systems.
The situation is more complicated when the number of point
perturbation is infinite. The spectral content of such models is
very rich and difficult to treat in the general setting; this is
why we restrict our attention here to the particular case
specified above, in which the periodicity allows us to employ the
Bloch-Floquet decomposition, with its specific features due to
presence of a homogeneous magnetic field that were first pointed
out by J.~Zak in his classical papers \cite{Z}.

On the other hand, the present work represents a generalization of
the results derived in \cite{G} for a two-dimensional
Schr\"odinger operator with a periodic point potential and a
homogeneous magnetic field perpendicular to the plane. Loosely
speaking we `add' the third dimension in the direction of the
magnetic field and allow the positions of the point potentials to
`spread' transversally preserving the periodicity -- this is what
we meant by a two-dimensional lattice in the first paragraph.

Our basic tool for spectral analysis is the Krein formula which
expresses the resolvent perturbation due to the point
interactions. It is important that in absence of the impurities
the Hamiltonian of the system allows a separation of the plane
variables from the transverse one. As a consequence, the free
resolvent kernel of the fiber operator (as well as all quantities
derived from it such as Bloch eigenfunctions, etc.) can be written
by means of explicitly given series. In this sense our model is
solvable, the only difference from those in the full space is that
there such quantities can be written in terms of elementary or
special functions.

In accordance with \cite{G} we adopt two more simplifying
assumptions: we suppose that the number of point interactions in
an elementary cell of the lattice is \emph{finite} and the flux
through the cell is \emph{rational}. Especially the second one is
important because one naturally expects in analogy with the
two-dimensional case that the spectral character can be
substantially different for an irrational flux; it is an
independent problem of its own interest which we are not going to
discuss here. On the other hand, following the setting of the
paper \cite{G} we study general point perturbations including the
situations when the particle can ``jump'' between different
impurity sites within a cell. The reason is that mathematically it
does not mean a lot of extra work; physically interesting case is
naturally that of point interactions defined by \emph{local}
boundary conditions, i.e. with the parameter matrix $A$ appearing
in the relation (\ref{bc}) being diagonal.

Let us review briefly the contents of the paper. We start in the
following section with analyzing the free Hamiltonian. Next, in
Section~\ref{sec mag}, we formulate the Bloch-Floquet theory for
the present case; the main result is so-called Landau-Zak
transformation which makes it possible to reduce the spectral
problem to investigation of suitable fiber operators. This is done
by means of Krein's formula in Section~\ref{sec perturb}. The last
three sections are devoted to successive analysis of the spectra
at three levels of complexity: \emph{(i)} an integer flux (in the
units of magnetic-flux quanta) through the elementary cell and a
single point interaction in the cell, \emph{(ii)} an integer flux
and a finite number of impurities in the elementary cell,
\emph{(iii)} and finally, the case of an arbitrary rational flux.


\setcounter{equation}{0}
\section{Free Hamiltonian}

Consider an infinite layer of a fixed width $d$, i.e. $\Sigma =
\R^2 \times [0,d]$. The coordinates we are going to use are $\vec
x=(x,x_3)$, where $x=(x_1,x_2)\in \R^2$ and $x_3 \in [0,d]$. The
layer is placed into a homogeneous magnetic field of intensity
$\vec B = (0,0,B)$. We will use the circular gauge, $\vec A =
{1\over 2} B (-x_2 ,x_1,0)$, and rational units, $\hbar=c=e=2m=1$.
A quantum particle confined to the layer is described by the free
magnetic Schr\"odinger operator in $L^2(\Sigma)$,
  \begin{equation}\label{free hamiltonian}
H_0 = (-i \vec\nabla - \vec A)^2,
  \end{equation}
with Dirichlet boundary conditions
  \begin{equation}\label{Dirichlet bc}
\psi(x,0)=\psi(x,d)=0\,, \quad x \in \R^2.
  \end{equation}
In the absence of an additional interaction the variables separate,
i.e. the operator $H_0$ can be decomposed into transverse modes,
  \begin{equation}\label{decomposition}
H_0 = \bigoplus_{n=1}^\infty\, h_n \otimes I\,, \quad h_n = \left(
-i {\partial \over \partial x_1} + {B \over 2} \, x_2 \right)^2 +
\left( -i {\partial \over
\partial x_2} - {B \over 2}\, x_1 \right)^2 + \left( {\pi n \over d}
\right)^2,
  \end{equation}
where $I$ is the unit operator in $L^2(0,d)$. The first two terms
at the r.h.s., in the following denoted as $h$, describe a
two-dimensional particle in the perpendicular homogeneous field in
the circular gauge. The third term represents the energy of the
$n$th transverse mode; the corresponding normalized eigenfunction
will be denoted as $\chi_n:\:  \chi_n(x_3)= \sqrt{2\over d}\, \sin
\left( {n\pi x_3 \over d} \right)$. The resolvent kernel of the
operator $h$ is well known \cite{DMM} to be
  \begin{eqnarray*}
\left( h - z \right)^{-1} (x,x') &\!=\!&  {1 \over 4\pi}\, \exp
\left( -i {B \over 2}\, x\wedge x' - {|B| \over 4}\, |x\!-\!x'|^2
\right)
\\ && \times\: \Gamma \left( {|B|\!-\!z \over 2 |B|} \right)
U \left( {|B|\!-\!z \over 2 |B|}, 1; {|B| \over 2}\, |x\!-\!x'|^2
\right)\,,
  \end{eqnarray*}
where $U$ is the irregular confluent hypergeometric function,
$\Gamma$ is the gamma function and $x\wedge x':= x_1 x'_2- x_2
x'_1$. For the sake of brevity we denote this kernel as
$G^{2D}_0(x,x';z)$. The decomposition (\ref{decomposition}) then
yields an explicit form for the resolvent kernel of the operator
(\ref{free hamiltonian}), namely
  \begin{eqnarray}\label{free resolvent}
G_0 (\vec x,\vec x';z) &\!\equiv\!& \left( H_0 - z \right)^{-1}
(x,x_3; x',x_3') \nonumber \\ &\!=\!& \sum_{n=1}^\infty G^{2D}_0
\left(x,x';z\!-\! \left( \pi n \over d \right)^2 \right)
\chi_n(x_3) \chi_n(x_3') \nonumber \\ &\!=\! & {1 \over 2 \pi d}\,
\exp \left( -i{B \over 2}\, x\wedge x' - {|B| \over 4}\,
|x\!-\!x'|^2 \right) \nonumber \\ &\times& \sum_{n=1}^\infty\:
\Gamma \left( {|B|\!-\!k_n^2(z) \over 2 |B|} \right) U \left(
{|B|\!-\!k_n^2(z) \over 2 |B|}, 1; {|B| \over 2}\, |x\!-\!x'|^2
\right) \nonumber
\\ &\times& \sin \left( {n\pi x_3 \over d} \right) \sin \left( {n\pi
x_3' \over d} \right),
  \end{eqnarray}
where $k_n:=\sqrt{z-(\pi n/ d)^2}$. Applying Fourier
transformation to the transverse part, ${\cal M}: \;
L^2([0,d])\mapsto \ell^2(\N)$ defined by
  \begin{equation}
({\cal M}f) (n)=\int_0^d dy \, \chi_n(y) f(y)\,, \quad n\in\N\,,
  \end{equation}
one can rewrite the free resolvent kernel as a matrix with the
elements
  \begin{equation}
G_0 (x,n; x',n' ;z) = \delta_{nn'} G_0^{2D} \left(x,x';z-\left(
{\pi n \over d}\right)^2 \right)\, .
  \end{equation}
Consequently, the spectrum of the free Hamiltonian consists of
Landau levels shifted by the energies of the transverse modes,
  \begin{equation}
\sigma \left( H_0 \right) = \sigma_{ess} \left( H_0 \right) =
\left\{ |B| (2 m +1) + \left( {\pi n \over d} \right)^2: \;
m,n\!-\!1 \in \N_0 \right\}.
  \end{equation}
The energy corresponding to the $l$th Landau level and $n$th
transverse mode will be denoted as $\varepsilon(l,n):=|B|(2 l +1)
+ ( \pi n /d)^2$. If the $|B|$ and $(\pi/d)^2$ are rationally
related it may happen that there exist more than one pair of
numbers $(l_i,n_i)$ giving the same value,
$\varepsilon(l_i,n_i)=z_0$. In such a case we will denote this set
of pairs as $J(z_0)$ and the number of these pairs as $|J(z_0)|$.


\setcounter{equation}{0}
\section{Magnetic translations}\label{sec mag}

Next we consider a lattice $\Gamma= \Lambda + K$, where $\Lambda$
is a lattice spanned by two independent vectors in $\R^2$, which
can be without loss of generality supposed to be $\vec
a=(a_1,0,0)$ and $\vec b=(b_1,b_2,0)$ with $b_2\ne 0$, and $K$ is
a set of $|K|$ points $\vec \kappa^i, \; i=1,\ldots,n$, from the
elementary cell $\{s a + t b: \; s,t\in[0,1)\} \times (0,d)$ of
the (interior of the) layer.

As usual with periodic systems, the first thing to do is to find
the appropriate representation of the translation group. In
presence of a magnetic field the argument shift must be composed
with multiplication by a suitable phase factor. Specifically, the
translation by a vector $\vec v =(v,0)$ acts in the state space
$L^2(\Sigma)$ of our system as follows,
   \begin{equation}
f(\vec x) \mapsto \exp (-\pi i \xi \, x\wedge v) f(\vec x \!-\!
\vec v)\,,
   \end{equation}
where $\xi$ is the number of the flux quanta of the field $\vec B$
through the unit area of the plane $\R^2$. Recall that the quantum
of magnetic flux is given by $2\pi \hbar c/|e|$, thus with the
chosen system of units we have $\xi=B/(2\pi)$. The phase factor at
the r.h.s. can be rewritten as $\exp \left( {1\over 2}(\vec B
\times \vec v).\vec x \right)$. Since $\vec B=(0,0,B)$ and we
consider translations in the plane, i.e. by a vector $\vec
v=(v_1,v_2,0)$, it is obvious that the third coordinate plays a
marginal role only in magnetic translations. This allows us to
follow closely the procedure used in \cite{G} to analyze spectral
properties of a particle confined to a plane, i.e. with the third
components of the vectors $\vec v$ and $\vec x$ absent.

Let us briefly summarize needed notions and facts. First we define
the group of discrete magnetic translation over the lattice
$\Lambda$,
   \begin{equation}
W(\xi,\Lambda) = \{ (\vec \lambda,\zeta): \; \vec
\lambda\in\Lambda, \; \zeta=\exp(\pi i \eta n), \; n\in\Z \}\,,
   \end{equation}
where $\eta=a_1 b_2 \xi= a\wedge b \, \xi$ is the number of flux
quanta of the field $\vec B$ through the elementary cell. In the
coordinates relative to the basis $\vec a$ and $\vec b$ the
multiplication in $W(\xi,\Lambda)$ has the form
   \begin{equation}
(\vec \lambda,\zeta) (\vec \lambda',\zeta') = \left(\vec \lambda +
\vec \lambda', \zeta \zeta' \exp (\pi i \eta(\lambda_a \lambda_b'
- \lambda_b \lambda_a') )\right)\,.
   \end{equation}
Notice that groups $W(\xi_i, \Lambda_i)$ corresponding to
different values of $\xi_i$ and different lattices $\Lambda_i$ but
having the same value of $\eta$ are isomorphic, hence we will
denote the group simply by $W_{\eta}$. Next we define the
representation $T$ of the group $W_\eta$ in the space
$L^2(\Sigma)$ as follows
   \begin{equation}
(T(\vec \lambda,\zeta)f)(\vec x) = \zeta \exp (-\pi i \xi \,
x\wedge \lambda) f(\vec x - \vec \lambda)\,.
   \end{equation}
Furthermore, replacing $\vec x\in \Sigma$ in the last formula by
vectors $\vec \gamma \in \Gamma$ we get a representation $D$ of
$W_\eta$ in the space $\ell^2(\Gamma)$.

If the flux $\eta$ is a rational number, then any unitary
representation of the group $W_\eta$ can be uniquely decomposed
into orthogonal sum of irreducible representations. Any `physical'
irreducible representations can be parameterized by a
point $p=(p_1,p_2)$ from the torus $T^2=[0,1)\times [0,1)$ -- see
\cite{OT}. If $\eta$ is an integer, then the group is Abelian and
the irreducible representations are one-dimensional, their
characters being given by
   \begin{equation}\label{represent int}
\chi((\vec \lambda,\zeta);p)=\zeta \exp \bigg(-2\pi i (\lambda_a
p_1 + \lambda_b p_2 +(N/2) \lambda_a \lambda_b)\bigg)\,.
   \end{equation}
For a general rational flux, $\eta=N/M$, the irreducible
representations of $W_\eta$ are generally $M$-dimensional. In
particular, the generators $(\vec a,1)$ and $(\vec b,1)$ are in
this case given by
   \begin{eqnarray}\label{represent rat}
\triangle((\vec a,1);p) &=& \mathrm{diag} \left[ e^{-2\pi i p_1},
e^{-2\pi i (p_1+ \eta)}, \ldots, e^{-2\pi i (p_1+ (M-1)\eta)}
\right]\,, \nonumber \\ \triangle((\vec b,1);p) &=& \left[
\begin{array}{ccc} 0 & I_{M-1} \\ e^{-2\pi i p_2} & 0
\end{array} \right] \,,
   \end{eqnarray}
where zeros in the second line are $(M\!-\!1)$-dimensional column
and row vectors, respectively. In order to obtain nonequivalent
representations, of course, we must restrict here $p$ to the torus
$T_\eta^2= [0,M^{-1})\times[0,1)$.

The decomposition into irreducible representations is accomplished
by the Landau-Zak transformation $\LL_\eta$. Recall that the
magnetic translations from $W_\eta$ does not affect the third
coordinate, or in the transverse-modes representation, they do
not affect the quantum number $n$. By a straightforward
modification of the two-dimensional formulae \cite{G}, we define
thus $\LL_\eta$ by
   \begin{eqnarray}\label{Landau-Zak transf}
\LL_\eta  &:&  L^2(\R^2) \otimes \ell^2(\N) \,\to \,
L^2(T^2_\eta)\otimes \C^M \otimes \C^N \otimes \ell^2(\N) \otimes
\ell^2(\N) \nonumber, \\ && (\LL_\eta f) (p,j,k,l,n) = N^{-{1
\over 2}} \sum_{m=-\infty}^{\infty} \exp \left(2\pi i m {p_2+k
\over N} \right)
\\ && \phantom{AAAAAAAAAAA} \times \int_{\R^2} d^2x f(x,n)
 \psi^*_0(x; p_1 +\eta j +m,l)\,, \nonumber \\
&& (\LL_\eta^{-1} \widetilde{f}) (x,n) = N^{-{1 \over 2}}
\sum_{j=0}^{M-1} \sum_{k=0}^{N-1} \sum_{l=0}^{\infty}
\sum_{m=-\infty}^{\infty} \int_0^{1/M} dp_1 \int_0^1 dp_2
\widetilde{f} (p,j,k,l,n) \nonumber \\ && \phantom{AAAAAAAAAAA}
\times \exp \left(-2\pi i m {p_2+k \over N} \right) \psi_0(x; p_1
+\eta j +m,l) \nonumber\,,
   \end{eqnarray}
where $\psi_0(x;q,l)$ with $q \in \R$ and $l=0,1,\ldots\,$, are
generalized eigenfunction of the operator $h$ (the `planar' part
of the Hamiltonian $H_0$) associated with the lattice $\Lambda$,
  \begin{eqnarray*}
\lefteqn{\psi_0(x;q,l) = \left( {b_2 \over \eta} \, {\pi^{3/2}
\over |B|^{3/2}}\, 2^{l+1} \, l!\right)^{-{1 \over 2}} \exp \bigg(
i\pi {b_1 \over a_1 \eta} \, q^2 \bigg) \exp \Bigg( 2\pi i {x_1
\over a_1} \bigg( {\eta \over b_2} \, {x_2 \over 2} +q \bigg)
\Bigg)} \\ && \phantom{AAAAA} \times \exp \Bigg( -|B| \bigg( x_2 +
{b_2 \over \eta}\, q \bigg)^2 \Bigg) H_l \bigg( |B|^{{1 \over 2}}
\bigg( x_2+ {b_2 \over \eta}\, q \bigg) \bigg)\,, \phantom{AAAAAA}
  \end{eqnarray*}
with $H_l$ being the $l$th-order Hermite polynomial. As in the
two-dimensional case, we get the following claims.
\begin{theorem}\label{thm: decomposition}
Assume that the flux is a rational number, $\eta=N/M$. Then the
Landau-Zak transformation (\ref{Landau-Zak transf}) decomposes the
representation $T$ of a magnetic translation $(\vec v,\zeta) \in
W_\eta$ into a direct integral of multiples of irreducible
representations $\triangle(\cdot)$, in other words
  \begin{equation}\label{represent decomposition}
\LL_\eta T(\vec v,\zeta) \LL_\eta^{-1} = \int_{T^2_\eta}^\oplus
d^2p\: \triangle (\vec v,\zeta;p) \otimes I_{\C^N} \otimes
I_{\ell^2(\N)} \otimes I_{\ell^2(\N)}.
  \end{equation}
In particular, if the flux $\eta$ is integer, then the irreducible
representations are one-dimensional and their characters are given
by the expression (\ref{represent int}).
\end{theorem}
The second important feature of the Landau-Zak transformation is
that it diagonalizes the free Hamiltonian in the space
$L^2(T^2_\eta)$, as well as any other operator
which exhibits the magnetic-translation symmetry:
\begin{theorem}\label{thm: operator decomposition}
Assume that a self-adjoint operator $H$ acting on $L^2(\Sigma)$ is
invariant w.r.t. the group $W_\eta$; then it can be decomposed
into a direct integral
  \begin{equation}
\widetilde{H} = \int_{T^2_\eta}^\oplus d^2p \, \widetilde{H}(p)\,,
  \end{equation}
where $\widetilde{H}=(\LL_\eta \otimes {\cal M}) H (\LL_\eta
\otimes {\cal M} )^{-1}$ and the fiber operator $\widetilde{H}(p)$
acts on the space $\C^M \otimes \C^N \otimes \ell^2(\N) \otimes
\ell^2(\N)$.
\end{theorem}


\setcounter{equation}{0}
\section{The perturbed Hamiltonian}\label{sec perturb}

Now let us consider perturbation of the free Hamiltonian $H_0$ by
point potentials placed at the points of the lattice $\Gamma$. To
define the perturbed operator $H_A$, where $A$ are the
coupling-parameter matrix introduced by the relation (\ref{bc})
below, we employ the standard technique based on theory of
self-adjoint extensions of symmetric operators. First, we pass to
the symmetric operator $S_\Gamma$ which is the restriction of
$H_0$ to the domain $D(S_\Gamma):=\{ f\in D(H_0): \; f(\vec
\gamma)=0, \; \vec \gamma \in\Gamma \}$, it is well defined in
view of the usual Sobolev embedding. The sought Hamiltonian $H_A$
is then an appropriate self-adjoint extension of the operator
$S_\Gamma$. There is a family of such extensions which can be
parametrized in different ways, the most traditional one is
the von Neumann method using unitary maps between deficiency
subspaces. For our purposes it is more suitable to employ the
mentioned self-adjoint operator $A$ acting in
$\ell^2(\Gamma)$; its properties will be specified later.

One of possible characterizations of the extensions starts from
the (generalized) boundary values. In this approach, one determines
the domain $D(H_A)$ as the set of all functions having prescribed
behavior in the vicinity of the point-potential sites $\vec \gamma
\in \Gamma$, namely
$$ \psi(\vec x)= L_0(\psi;\vec \gamma) {1+ i \vec A (\vec \gamma).
(\vec x -\vec \gamma)\over |\vec x - \vec \gamma|}
+ L_1(\psi;\vec \gamma)+\OO(|\vec x -\vec \gamma|), $$
where the vectors $L_j(\psi):= \{ L_j(\psi;\vec \gamma):\:
\vec\gamma\in \Gamma\, \}\,, j=0,1\,$, are related by
   \begin{equation}\label{bc}
L_1(\psi)= - 4\pi A L_0(\psi)\,.
   \end{equation}
This is a generalization of the usual point-interaction definition
-- see \cite{AGHH} -- the latter corresponds to a diagonal $A$ and
represents the physically most interesting situation, at least
from the viewpoint of modeling semiconductor layers with
impurities, as we pointed out in the introduction.

Spectral properties of the Hamiltonian can be found by means of
its resolvent. Krein's formula gives us its kernel, i.e. the
Green function of $H_A$ for a fixed operator $A$,
  \begin{equation}\label{Krein}
G(\vec x, \vec x';z) = G_0(\vec x, \vec x';z) -
\sum_{\gamma,\gamma'\in \Gamma} [Q(z)+A]^{-1} (\vec \gamma,
\vec\gamma') G_0(\vec x, \vec\gamma;z) G_0(\vec\gamma',\vec
x';z)\,.
  \end{equation}
with $Q(z)$ given by the relation (\ref{def Q}) below. If we use
the transverse-mode representation of the Hamiltonian in the space
$L^2(\R^2)\otimes \ell^2 (\N)$ we get
  \begin{eqnarray}
G(x,n; x',n';z) &=& \delta_{nn'} G_0^{2D} \left(x, x'; z- \left(
{\pi n\over d} \right)^2 \right)
\\ &&
- \sum_{\vec \gamma,\vec \gamma'\in \Gamma} [Q(z)+A]^{-1} (\vec
\gamma, \vec\gamma') G_0^{2D} \left(x, \gamma; z - \left( {\pi n
\over d} \right)^2 \right)  \nonumber \\ && \times G_0^{2D}
\left(\gamma', x'; z - \left( {\pi n' \over d} \right)^2 \right)
\chi_n(\gamma_3) \chi_{n'}(\gamma_3') \nonumber.
  \end{eqnarray}
Since the Krein's formula (\ref{Krein}) was originally meant for a
finite number of point potentials and the set $\Gamma$ is
infinite, we have to be a bit more precise. One possible way how
to define the Hamiltonian $H_A$ is via strong resolvent limit of
the family of ``restricted'' operators $H_A (\bar{\Gamma})$ which
are point-interaction Hamiltonians referring to finite subsets
$\bar{\Gamma} \subset \Gamma$, an example for diagonal $A$ and
$B=0$ can be found in \cite[III.1.1]{AGHH}. But the limit is not
necessary; the existence of the operator $H_A$ and the generalized
Krein formula have been already proven for a larger class of
operators, -- see, e.g., \cite{GMC} or \cite{P}.

The matrix $Q$ in (\ref{Krein}) is given by
  \begin{equation}\label{def Q}
Q(\vec \gamma, \vec \gamma';z) = \left\{ \begin{array}{ll}
G_0(\vec \gamma, \vec \gamma';z) \qquad & \vec\gamma \neq
\vec\gamma' \\ Q_0( \gamma_3;z) \qquad & \vec\gamma = \vec\gamma'
\end{array} \right. ,
  \end{equation}
where $Q_0(\gamma_3;z)$ is the regularized Green function
(stripped off the pole singularity) which is defined by
$\lim_{|\vec x - \vec\gamma| \to 0}(G_0(\vec x, \vec\gamma;z)- {1
\over 4\pi} |\vec x - \vec\gamma|^{-1})$. We know from \cite{EN}
how it looks like,
  \begin{eqnarray}\label{xi B}
Q_0 (\gamma_3;z) &\!=\!& {1 \over 2\pi d} \sum_{n=1}^\infty
\left[\, \ln \left((\pi n)^2 \over 2 |B| d^2\right) - \psi \left(
{|B|-z+ \left( {\pi n \over d} \right)^2 \over 2|B|} \right)
\right] \sin^2 \left( {\pi n \gamma_3 \over d} \right) \nonumber
\\&& +\, {1 \over 4\pi d}\, \left[ C_E + \psi \left( {\gamma_3 \over d}
\right)+ {\pi \over 2} \cot \left( {\pi \gamma_3 \over d} \right)
\right]\,.
  \end{eqnarray}
In case when two point potentials are arranged vertically, i.e.
$\gamma=\gamma'$ and $\gamma_3\neq\gamma_3'$, the corresponding
element of $Q$ is well defined but the expression (\ref{free
resolvent}) makes no sense and has to be recast into the form
  \begin{eqnarray}
Q (\vec \gamma,\vec \gamma';z) &=& {1 \over 2\pi d}
\sum_{n=1}^\infty \left[\, \ln \left((\pi n)^2 \over 2 |B|
d^2\right) - \psi \left( {|B|-z+ \left( {\pi n \over d} \right)^2
\over 2|B|} \right) \right] \nonumber \\ && \phantom{AAAA} \times
\sin \left( {\pi n \gamma_3 \over d} \right) \sin \left( {\pi n
\gamma_3' \over d} \right) \nonumber
\\ && +\, {1 \over 4\pi d}\, \left[ C_E + \psi \left(
{\gamma_3+\gamma_3'
 \over 2d} \right)+ {\pi \over 2} \cot \left( {\pi (\gamma_3+\gamma_3')
\over 2d} \right) \right] \nonumber \\ && -\, {1 \over 4\pi d}\,
\left[ C_E + \psi \left( {|\gamma_3-\gamma_3'| \over 2d} \right)+
{\pi \over 2} \cot \left( {\pi |\gamma_3-\gamma_3'| \over 2d}
\right)\right]\,. \nonumber
  \end{eqnarray}
For the sake of brevity it is useful to rewrite the Krein's
formula in a compact form,
  \begin{equation}\label{compact Krein}
R_A(z)=R_0(z)-\Gamma_z [Q(z)+A]^{-1} \Gamma_z^*,
  \end{equation}
where $\Gamma_z$ is a map from $\ell^2(\Gamma)$ to $L^2(\R^2)
\otimes \ell^2(\N)$ defined by
  \begin{equation}
(\Gamma_z \phi) (x,n) := \sum_{\vec \gamma \in \Gamma} \phi (\vec
\gamma) G_0^{2D} \left(x, \gamma; z- \left( {\pi n \over d}
\right)^2 \right) \chi_n(\gamma_3).
  \end{equation}
The operator $H_0$ is invariant under magnetic translations from
the group $W_\eta$. One can easily check that the free resolvent
kernel satisfies the relation
   \begin{equation}
G_0(\vec x -\vec \lambda, \vec x' -\vec \lambda;z) = \exp (\pi i
\xi \, (x-x')\wedge \lambda) G_0(\vec x, \vec x';z).
   \end{equation}
Our aim is to study the situation when the operator $H_A$ is also
$W_\eta$-invariant, which is the case when the operator $A$ is
assumed to satisfy the same condition. Notice that this is
trivially satisfied if $A$ is a diagonal matrix. We will assume
only that $A$ is a self-adjoint operator invariant w.r.t.
$W_\eta$, and furthermore, that it is bounded and there exist two
positive constants $c_1$ and $c_2$ such that
   \begin{equation}\label{estimate A}
|A(\vec \gamma,\vec \gamma')| \leq c_1 \exp (-c_2
|\gamma-\gamma'|) \quad {\rm for \; all} \; \vec\gamma, \gamma'
\in \Gamma\,.
   \end{equation}
The last condition means a restriction on the non-locality we have
allowed mathematically: it means that the probability of particle
hoping between two points of the lattice $\Gamma$ decays
exponentially with their distance.

A similar estimate is valid for the operator $Q$ -- for any given
$z\notin\sigma(H_0)$ there exist two positive constants $c_3$ and
$c_4$ such that
   \begin{equation}\label{estimate Q}
|Q(\vec \gamma,\vec \gamma';z)| \leq c_3 \exp (-c_4 |\gamma-
\gamma'|) \quad {\rm for \; all} \; \vec\gamma, \gamma' \in \Gamma
\, ,
   \end{equation}
as it follows from the definition (\ref{def Q}) of the function
$Q$ and the free resolvent kernel (\ref{free resolvent}). The
infinite sum contained in the second formula converges because the
term $\Gamma(u)U(u,1;s)$ can be written for large positive $u$ as
$2K_0(2\sqrt{u s})$ (see \cite[13.3.3]{AS}) and the Macdonald
function $K_0$ decays exponentially for large argument. Using the
asymptotics of the function $U(u,1;s)$ we conclude that the sum
grows with $|\gamma|$ at most as a polynomial; hence the
exponential term $\exp (-{1\over 4}|B| |\gamma|^2)$ is sufficient
to yield the estimate.

If $z$ approaches a point from the spectrum of the free
Hamiltonian $H_0$ the elements of $Q$ may diverge -- cf.~\cite{EN}
-- in other words the functions $Q(\vec \gamma, \vec
\gamma';\cdot)$ are in general meromorphic. We summarize the above
discussion in the following theorem:
\begin{theorem}\label{thm: s-a extension}
Suppose that an operator $A$ acting on $\ell^2(\Gamma)$ is
self-adjoint and $W_\eta$-invariant. Then there is
exactly one self-adjoint extension $H_A$ of the operator
$S_\Gamma$ with Green function given by
  $$
G(\vec x, \vec x';z) = G_0(\vec x, \vec x';z) -
\sum_{\gamma,\gamma'\in \Gamma} [Q(z)+A]^{-1} (\vec \gamma,
\vec\gamma') G_0(\vec x, \vec\gamma;z) G_0(\vec\gamma',\vec
x';z)\,,
  $$
where the operator $Q(z)$ is defined by the
relations (\ref{def Q}) and (\ref{xi B}). As an operator in
$\ell^2(\Gamma)$, i.e. an infinite matrix, $Q(\vec \gamma,\vec
\gamma';z)$ satisfies the estimate (\ref{estimate Q}) for some
$c_3,c_4>0$ and it is $W_\eta$-invariant. As a function of $z$,
$Q$ is meromorphic and all its poles belong to the spectrum
$\sigma(H_0)$. Finally, the operator $H_A$ is also
$W_\eta$-invariant.
\end{theorem}


\setcounter{equation}{0}
\section{The case of integral flux and a monoatomic crystal}

We begin with the simplest case assuming that the flux $\eta$
through the elementary cell is integral, $\eta=N\geq 1$, and that
this cell contains only one potential placed at $\vec
\kappa=(0,0,\kappa_3)$; in other words in this section we have
$\Gamma=\Lambda+\{\vec\kappa\}$.

We already know from Thm.~\ref{thm: decomposition} that to
diagonalize the representation $T$ acting in $L^2(\Sigma)$ we have
to employ the Landau-Zak transformation. Since $M=1$ in the
present case we drop the parameter $j$. Diagonalization of the
representation $D$ acting in $\ell^2(\Gamma)$ is achieved by the
Fourier transformation ${\cal F}_\eta: \; \ell^2(\Lambda) \mapsto
L^2(T^2)$ which is defined by
   \begin{equation} \label{Four}
({\cal F}_\eta \phi) (p) = \sum_{\vec \lambda \in \Lambda}
\phi(\vec \lambda+\vec \kappa) e_{\lambda}(p)
   \end{equation}
with the basis $e_{\lambda}(p):= \exp (-2\pi i (\lambda_a p_1
+\lambda_b p_2 +N \lambda_a \lambda_b /2))$. It reduces the action
of the magnetic translation to multiplication by the function
$e_\lambda(p)$. In view of the $W_\eta$-invariance of $G_0$, the
transformed function $\widetilde{Q}$ equals
   \begin{equation}\label{tilde Q}
\widetilde{Q}(p;z) = ({\cal F}_\eta Q(z) {\cal F}_\eta^{-1}) (p) =
\sum_{\vec \lambda \in \Lambda} Q(\vec \lambda,\vec\kappa;z)
e_{\lambda}(p).
   \end{equation}
With the exponential estimate (\ref{estimate Q}) of the function
$Q$ in mind, we infer that the function $\widetilde{Q}$ is well
defined. It is also meromorphic in $z$ with simple poles, which
lie in $\sigma(H_0)$. Both the functions
$\widetilde{Q}(p;z)$ and $\widetilde{A}(p)$ are real-analytic with
respect to $p_1,p_2 \in \R$.

Our goal now is to find the $\LL_{\eta}$-transformation of the
Green function given by Krein's formula (\ref{compact Krein}).
To this aim we denote
   \begin{eqnarray*}
\widetilde{R}_A(z) &=& \LL_\eta R_A(z) \LL_\eta^{-1}\,, \\
\widetilde{R}_0(z) &=& \LL_\eta R_0(z) \LL_\eta^{-1}\,, \\
\widetilde{\Gamma}_z &=& \LL_\eta \Gamma_z {\cal F}_\eta^{-1}\,.
   \end{eqnarray*}
After a straightforward computation we arrive at the formula
  \begin{eqnarray}\label{Green function}
\lefteqn{ G(p;k,l,n;k',l',n';z) = \delta_{kk'}\, \delta_{ll'}\,
\delta_{nn'} \,{1 \over \varepsilon(l,n)-z} } \\ && - \,
[\widetilde{Q}(p;z)+\widetilde{A}(p)]^{-1} {\widetilde{\delta}_0
(p;k,l) \over \varepsilon (l,n)-z} \; {\widetilde{\delta}_0^*
(p;k',l') \over \varepsilon (l',n')-z} \chi_n(\kappa_3)\,
\chi_{n'}(\kappa_3) \nonumber\,,
  \end{eqnarray}
where $\widetilde{\delta}_0 (p;k,l)$ is the ${\cal
L}_\eta$-transformed delta-function in $\R^2$,
  \begin{equation}
\widetilde{\delta}_\gamma (p;k,l) = N^{-1/2}
\sum_{m=-\infty}^\infty \exp \left(2\pi i m {p_2+k \over N}\right)
\psi_0^* (\gamma,p_1+m,l)\,;
  \end{equation}
we have used here the relation
  \begin{equation}\label{shifted delta}
\widetilde{\delta}_{\lambda+\gamma} (p;k,l) =\exp(\pi i \xi \,
\gamma\wedge \lambda) \, e_\lambda(p) \,
\widetilde{\delta}_{\gamma} (p;k,l)\,,
  \end{equation}
which follows from the $W_\eta$-invariance of the free resolvent
kernel and the fact that $\widetilde{\delta}_\gamma (p;k,l) =
(|B|(2l+1)-z) (\LL_\eta G_0^{2D} (\cdot,\gamma;z)) (p;k,l)$.
Recall that although the Landau-Zak transformation $\LL_\eta$
acts on $L^2(\R^2)\otimes \ell^2(\N)$, it can be viewed as
two-dimensional since it does not affect the transverse modes.

To perform the spectral analysis we need to know the behavior of
the function $\widetilde{Q}(p;z)$ for fixed $p$ and real $z$. It
is convenient to treat this problem separately in each of the
(infinitely many) intervals corresponding to gaps in the free
Hamiltonian spectrum. To this end, we denote the points of $\sigma
(H_0)$ arranged in the ascending order by $\varepsilon_i$,
$i=0,1,\ldots\,$. The function $\widetilde{Q}(p,\cdot)$ diverges
at a chosen point $\varepsilon_i$ if and only if there exists at
least one pair of integer numbers $(l,n) \in J(\varepsilon_i)$
such that $\chi_n(\kappa_3)\neq 0$ and $p$ belongs to $U_l=\{ p
\in T^2: \; \widetilde{\delta}_0 (p;\cdot,l)\neq 0 \}$, where the
expression $\widetilde{\delta}_0 (p;\cdot,l)$ stands for an
$N$-dimensional vector. By \cite{KL} the residues in Krein's
formula (\ref{compact Krein}) are given by $dQ/dz=\Gamma^*_{z^*}
\Gamma_z$; the transformed form of this relation reads
  \begin{equation}\label{derivative Q}
{\partial \widetilde{Q} \over \partial z}(p;z) = \sum_{l=0}^\infty
\sum_{n=1}^\infty {1 \over |\varepsilon(l,n)-z|^2} \;
\chi_n^2(\kappa_3) \sum_{k=0}^{N-1} |\widetilde{\delta}_0
(p;k,l)|^2 \nonumber.
  \end{equation}
For notational convenience, we also put $\varepsilon_{-1}=-\infty$
and $U_{-1}=T^2$. It is clear from (\ref{derivative Q}) that the
function $\widetilde{Q}(p,\cdot)$ is monotonously increasing in
each of the intervals $(\varepsilon_{i-1}, \varepsilon_i)$.

The asymptotic behaviour of $\widetilde{Q}(p,z)$ for large
negative $z$ is governed by the function $Q_0(\kappa_3;z)$, while
the contribution of the rest of the series in expression
(\ref{tilde Q}) for $\widetilde{Q}$ is bounded, its convergence
being ensured by the estimate (\ref{estimate Q}). By \cite{EN}
the divergent term has the expansion
  \begin{equation}\label{nonmagnetic asymp}
Q_0(\kappa_3;z)= - {\sqrt{-z} \over 4\pi} + \OO(1) \qquad {\rm as}
\qquad z \to -\infty\,.
  \end{equation}
Now we are ready to analyze the spectrum of the fibre operator
$\widetilde{H}_A(p)$. The first thing we would like to know is
whether the Landau levels of the free system stay in the spectrum
$\sigma(\widetilde{H}_A(p))$. By examining the residues of the
Green function (\ref{Green function}) at $z=\varepsilon_i \in
\sigma(H_0)$, $i=0,1,\ldots\,$, it is straightforward find the
multiplicity of $\varepsilon_i$ in $\sigma(\widetilde{H}_A(p))$
and the corresponding eigenspaces.
\begin{lemma}\label{lemma: multiplicity 1}
Assume that the flux $\eta$ equals an integer number $N$ and the
lattice $\Gamma$ is `monoatomic'. Fix a point $\varepsilon_i$ from
$\sigma(H_0)$ and $p \in T^2$, then one of the following three
situations occurs: \\ [1mm]
(i) if there is at least one pair $(l,n)\in J(\varepsilon_i)$
satisfying $\widetilde{\delta}_0 (p;\cdot,l) \neq 0$ and
$\chi_n(\kappa_3)\neq 0$, the multiplicity $d(p;\varepsilon_i)$
equals $N |J(\varepsilon_i)|-1$ and the eigenspace is the
orthogonal complement of the vector $\omega(p;\varepsilon_i)$ in
the space $\Omega(p;\varepsilon_i)$, where
  \begin{eqnarray}\label{eigenspace}
\omega(p;\varepsilon_i) &=& \Big( \widetilde{\delta}_0 (p;k,l)
\chi_n (\kappa_3) \sum_{(l_0,n_0)\in J(\varepsilon_i)}
\delta_{ll_0} \delta_{nn_o} \Big)_{k,l,n}\,, \nonumber \\
\Omega(p;\varepsilon_i) &=& \C^N \otimes \Big( \bigoplus_{(l,n)\in
J(\varepsilon_i)} (e_l \otimes \chi_n) \Big)\,,
  \end{eqnarray}\\
(ii) if $\chi_n(\kappa_3) \widetilde{\delta}_0 (p;\cdot,l)$ is a
zero vector for all indices $(l,n)\in J(\varepsilon_i)$ and
$\widetilde{Q}(p;\varepsilon_i)+\widetilde{A}(p)\neq 0$, then
$d(p;\varepsilon_i)=N |J(\varepsilon_i)|$ and
$\Omega(p;\varepsilon_i)$ is the eigenspace,\\ [1mm]
(iii) if $\chi_n(\kappa_3) \widetilde{\delta}_0 (p;\cdot,l)$ is a
zero vector for all indices $(l,n)\in J(\varepsilon_i)$ and
$\widetilde{Q}(p;\varepsilon_i)+\widetilde{A}(p)=0$, then
$d(p;\varepsilon_i)=N |J(\varepsilon_i)|+1$. In this case the
eigenspace is the linear hull of $\Omega(p;\varepsilon_i)$ and the
vector $\Big({\widetilde{\delta}_0 (p;k,l) \chi_n (\kappa_3) \over
\varepsilon(l,n)-\varepsilon_i} \Big)_{k,l,n}$, where we put
$0/0:=0$ for elements with $(l,n)\in J(\varepsilon_i)$.
\end{lemma}

This answers the question what happens with the free Hamiltonian
spectrum under influence of the perturbation. However, the
spectrum $\sigma(\widetilde{H}_A(p))$ contains not only the
original modified Landau levels but also additional points due to
the presence of the point potential. From the properties of
$\widetilde{Q}(p,\cdot)$ it is obvious that there is exactly one
solution $E_{i_k}(p)$ of the implicit equation
  \begin{equation}\label{implicit eq}
\widetilde{Q}(p;E)+\widetilde{A}(p)=0
  \end{equation}
in any interval $(\varepsilon_{i_{k-1}(p)},
\varepsilon_{i_k(p)})$, where
$(\varepsilon_{i_k(p)})_{k=-1}^\infty$ is a subsequence of all
points from $(\varepsilon_i)_{i=-1}^{\infty}$ at which the
function $\widetilde{Q}(p;\cdot)$ diverges. If $E$ does not belong
to $\sigma(H_0)$, the said solution is a nondegenarate eigenvalue
of $\widetilde{H}_A(p)$ with the unnormalized eigenvector $\left(
{\widetilde{\delta}_0 (p;k,l) \over \varepsilon(l,n) - E(p)}\,
\chi_n(\kappa_3) \right)_{k,l,n}$. This vector is non-trivial,
which follows from the inequality (\ref{positivity}) derived below
and applied to the derivative $\partial \widetilde{Q} /
\partial z$ given in the monoatomic case by (\ref{derivative Q}).

Let us summarize the effect of the periodic point potential on the
spectrum $\sigma(\widetilde{H}_A(p))$ at a fixed point $p \in
T^2$. The modified Landau level $\varepsilon_i$ in the spectrum of
the free Hamiltonian $H_0$ has the multiplicity equal to $N
|J(\varepsilon_i)|$. In the perturbed spectrum
$\sigma(\widetilde{H}_A(p))$ the generic situation is the case (i)
of the above lemma, when the function $\widetilde{Q}(p;\cdot)$
diverges at $\varepsilon_i$. Then an eigenvalue splits off this
level moving down with the increasing $\widetilde{A}(p)$ towards
the neighbouring lower modified Landau levels. In particular, for
$N |J(\varepsilon_i)|=1$ the perturbation removes in this way the
level $\varepsilon_i$ from the spectrum $\sigma
(\widetilde{H}_A(p))$ entirely. On the other hand, if
$\widetilde{Q}(p;\cdot)$ does not diverge at $\varepsilon_i$, the
multiplicity remains the same or it can be enlarged by one; the
latter happens when an eigenvalue coming from a higher modified
Landau level reaches $\varepsilon_i$ for a particular
$\widetilde{A}(p)$. These two situations correspond, of course, to
cases (ii) and (iii) of Lemma~\ref{lemma: multiplicity 1},
respectively.

After we have found the spectrum of the fibre operator for a fixed
quasimomentum, we can proceed to analysis of the spectrum of the
full operator
  \begin{equation} \label{dirintegral}
H_A \simeq \widetilde{H}_A = \int_{T^2}^\oplus \widetilde{H}_A (p)
\, dp\,.
  \end{equation}
Since the functions appearing in the equation (\ref{implicit eq})
are real-analytic the same is true for its solution. Fix an
interval $(\varepsilon_{i-1},\varepsilon_i)$ such that there are
two pairs of indices $(l_1,n_1) \in J(\varepsilon_{i-1})$ and
$(l_2,n_2) \in J(\varepsilon_i)$ with $\chi_{n_1}(\kappa_3) \neq
0$ and $\chi_{n_2} (\kappa_3) \neq 0$, then by the above
discussion there is a real-analytic function $E_i(\cdot)$ defined
on a set $U_{l_1} \cap U_{l_2}$ with the range in
$(\varepsilon_{i-1},\varepsilon_i)$. Moreover,
$\widetilde{A}(\cdot)$ is bounded as a continuous function on a
compact set. Combining this observation with  the asymptotics
(\ref{nonmagnetic asymp}) for $i=0$, we see that also the range of
$E_i(\cdot)$ is a bounded interval. Since
$\widetilde{\delta}_0(\cdot;k,l)$ is an analytic function and $U_l
\neq \emptyset$ for all $l \in \N \cup \{0\}$, the domains $U_l$
are dense open sets of full measure, so any intersection $U_{l_1}
\cap U_{l_2}$ is also an open set of full measure. Hence the
function $E_i(\cdot)$ extends by continuity to the entire torus
$T^2$ and its range lies in the interval $[\varepsilon_{i-1},
\varepsilon_i]$, having a finite lower bound for $i=0$.

A modification is needed in case of an `orphan' modified Landau
level, i.e. a point $\varepsilon_{i'}$ for which there is no pair
of indices $(l,n) \in J(\varepsilon_{i'})$ satisfying
$\chi_n(\kappa_3) \neq 0$. It is obvious that this cannot be a
pole of $\widetilde{Q}(p;\cdot)$ for any $p$, and consequently,
the implicit equation (\ref{implicit eq}) may have no solution in
one or both of the intervals $(\varepsilon_{i'-1},
\varepsilon_{i'})$ and $(\varepsilon_{i'}, \varepsilon_{i'+1})$.
Instead we have to consider in this case the joint interval
amended with the common endpoint. Then there is a unique solution
$E_{i'+1}(p)$  of (\ref{implicit eq}) on the interval
$(\varepsilon_{i'-1},\varepsilon_{i'+1})$, provided its endpoints
belong to the `regular' class considered above, and the dispersion
function $E_{i'}(p)$ is excluded from further consideration; the
argument easily extends to the situation with two or more
neighbouring `orphan' points. Note that the number of `non-orphan'
levels is infinite.

Eliminating the `orphan' points $\varepsilon_{i'}$ from
$(\varepsilon_{i})_{i=0}^\infty$ we obtain a subsequence
$(\varepsilon_{i_k})_{k=0}^\infty$; we add conventionally
$\varepsilon_{-1}=-\infty$ as its first term. For each interval
$(\varepsilon_{i_{k-1}}, \varepsilon_{i_k})$ we have then a unique
dispersion function $E_{i_k}(p)$ defined on an open set of full
measure as a solution to (\ref{implicit eq}) and  extended by
continuity to the entire torus $T^2$.
  \begin{lemma}\label{lemma: exten E 1}
The function $E_{i_k}(\cdot)$ defined above has following
properties:\\ [1mm]
(i) if $\varepsilon_{i_{k-1}}<E_{i_k}(p)<\varepsilon_{i_k}$, then
$E_{i_k}(p)$ is the unique solution to the implicit equation
(\ref{implicit eq}) in $(\varepsilon_{i_{k-1}},
\varepsilon_{i_k})$,\\ [1mm]
(ii) if $\varepsilon_{i_k}$ is a pole of the function
$\widetilde{Q}(p;\cdot)$, then $E_{i_k}(p)<\varepsilon_{i_k}<
E_{i_{k+1}}(p)$,\\ [1mm]
(iii) if $\varepsilon_{i_k}$ is not a pole of the function
$\widetilde{Q}(p;\cdot)$, then
  \begin{eqnarray*}
\widetilde{Q}(p;\varepsilon_{i_k})+\widetilde{A}(p)<0
&\Rightarrow& E_{i_k}(p)=\varepsilon_{i_k}<E_{i_{k+1}}(p)\,, \\
\widetilde{Q}(p;\varepsilon_{i_k})+\widetilde{A}(p)>0
&\Rightarrow& E_{i_k}(p)<\varepsilon_{i_k}=E_{i_{k+1}}(p)\,, \\
\widetilde{Q}(p;\varepsilon_{i_k})+\widetilde{A}(p)=0
&\Rightarrow& E_{i_k}(p)=\varepsilon_{i_k}=E_{i_{k+1}}(p)\,.
  \end{eqnarray*}
  \end{lemma}
{\sl Proof:} The proof is similar to the one for point
interactions on the plane \cite{G}. By definition of the sequence
$\{\varepsilon_{i_k}\}_{k=-1}^\infty$ there are pairs of indices
$(l_1,n_1) \in J(\varepsilon_{i_{k-1}})$ and $(l_2,n_2) \in
J(\varepsilon_{i_k})$ such that $\chi_{n_1}(\kappa_3) \neq 0$ and
$\chi_{n_2}(\kappa_3) \neq 0$. The union $U_{l_1} \cap U_{l_2}$
has a full measure in $T^2$, so the extension of $E_{i_k}(\cdot)$
to the whole $T^2$ by continuity (using a sequence
$\{p_n\}_{n=1}^\infty \in U_{l_1} \cap U_{l_2}$ tending to $p\in
T^2$) is well defined. Assume first that $E_{i_k}(p)$ does not
coincide with the endpoints $\varepsilon_{i_{k-1}}$ and
$\varepsilon_{i_k}$ of the interval. From the condition $\mu
(p_n;E_{i_k}(p_n)) := \widetilde{Q}(p_n;E_{i_k}(p_n))+
\widetilde{A}(p_n)=0$ and the joint continuity of
$\mu(\cdot;\cdot)$ in the a neighbourhood of the point
$(p,E_{i_k}(p))$ we infer that $\mu(p;E_{i_k}(p))=0$ for any $p\in
T^2$; then the claim (i) follows from the monotonicity of the
function $\mu(p;\cdot)$.

The point $E_{i_k}(p)$ defined as above cannot be a pole of the
function $\mu(p;\cdot)$. To see this, consider the function $z
\mapsto \beta(p;z):= \mu(p;z) (z-\varepsilon_{i_{k-1}})
(z-\varepsilon_{i_k})$, which is analytic in an interval
$(\varepsilon_{i_{k-1}}-\rho, \varepsilon_{i_k} +\rho)$ with a
small enough $\rho>0$. Using the continuity again we find that
$\beta(p;E_{i_k}(p))=0$ for any $p\in T^2$, thus a pole at
$E_{i_k}(p)$ is excluded. This proves the claim (ii).

To prove the
last statement of the lemma, we need two auxiliary results:
  \begin{eqnarray}\label{auxil ineq}
E_{i_k}(p)=\varepsilon_{i_k} &\Rightarrow& \lim_{z \to E_{i_k}(p)}
\mu(p;z) \leq 0\,, \nonumber \\ E_{i_k}(p)=\varepsilon_{i_{k-1}}
&\Rightarrow& \lim_{z \to E_{i_k}(p)} \mu(p;z) \geq 0 \,.
  \end{eqnarray}
Let us check the first relation. Assume that $E_{i_k}(p)=
\varepsilon_{i_k}$ but the limit is strictly positive, then there
are $E_0 \in (\varepsilon_{i_{k-1}}, \varepsilon_{i_k})$ with $\mu
(p; E_0)>0$ and $p_{\tilde{n}}$ such that $\mu (p_{\tilde{n}};
E_0)>0$ and $E_0 <E_{i_k}(p_{\tilde{n}})$. However, the
monotonicity of $\mu(p;\cdot)$ leads then to a contradiction, $0<
\mu (p_{\tilde{n}}; E_0)< \mu (p_{\tilde{n}};
E_{i_k}(p_{\tilde{n}})) =0$. Assume now that $\varepsilon_{i_k}$
is not a pole of $\mu(p;\cdot)$ and $\mu(p; \varepsilon_{i_k})<0$.
By (\ref{auxil ineq}) we would have $\varepsilon_{i_k}
<E_{i_{k+1}}(p)$. There is an integer number $n_0$ such that
$\mu(p_n;\varepsilon_{i_k})<0$ holds for all $n>n_0$, hence
$\varepsilon_{i_k}<E_{i_k}(p_n)$ for all $n>n_0$, and
consequently, $\varepsilon_{i_k} \leq E_{i_k}(p)$. The case of the
opposite inequality $\mu(p;\varepsilon_{i_k})>0$ is treated in a
similar way.

Finally, let us consider the last case when $\varepsilon_{i_k}$ is
not a pole of $\mu(p;\cdot)$ and $\mu(p;\varepsilon_{i_k})=0$. We
have to exclude both the strict inequalities $E_{i_k}(p)
<\varepsilon_{i_k}$ and $\varepsilon_{i_k} <E_{i_{k+1}}(p)$.
Assume, for instance, that the first one of them holds. From
(\ref{auxil ineq}) and the proof of the claim (i) we know that
$\lim_{z \to E_{i_k}(p)} \mu(p;z) \geq 0$. There exists $E_1$
satisfying $E_{i_k}(p)< E_1< \varepsilon_{i_k}$ and we arrive at a
contradiction, $0 \leq \mu(p;E_{i_k}(p)) < \mu(p;E_1) \leq
\mu(p;\varepsilon_{i_k})=0$.

To finish the proof, we must check that the definition of
$E_{i_k}(p)$ is independent of the choice of the approximating
sequence $(p_n)_{n=1}^\infty$. Consider another sequence
$(p'_n)_{n=1}^\infty$ converging to $p$ and denote the limit of
$E_{i_k}(p'_n)$ by $E'_{i_k}(p)$. Assume $E_{i_k}(p)
<E'_{i_k}(p)$, then by the claim (i) it is necessary that at least
one of these points coincides with one of the endpoints of the
interval $(\varepsilon_{i_{k-1}}, \varepsilon_{i_k})$. Using
(\ref{auxil ineq}) we arrive at the relations
$$ \lim_{z \to E_{i_k}(p)} \mu(p;z) \geq 0\,, \qquad \lim_{z \to
E'_{i_k}(p)} \mu(p;z) \leq 0\,. $$
Choosing points $E_1,\,E_2$ such that $E_{i_k}(p)< E_1< E_2<
E'_{i_k}(p)$, we get a contra\-diction, $0 \leq \mu(p;E_{i_k}(p))
\leq \mu(p;E_1)< \mu(p;E_2) \leq \mu(p;E'_{i_k}(p)) \leq 0$. $\:$
\QED \vspace{1em}

Combining the above results with with the direct-integral
decomposition (\ref{dirintegral}) we arrive finally at the sought
description of the spectrum $\sigma(H_A)$.
  \begin{theorem}\label{thm: spectrum 1}
Suppose that the flux $\eta$ is integer, $\eta=N$, and the
elementary cell contains one point potential. Then $\sigma(H_A)$
consists of two parts:\\ [1mm]
(i) The first one is the union of spectral bands denoted by $J_k$,
$k=0,1,\ldots$, where $J_k$ is the range of the function
$E_{i_k}(\cdot)$ over the torus $T^2$, with $E_{i_k}(p)$ defined
by the implicit equation (\ref{implicit eq}). Each band $J_k$ lies
within one interval $[\varepsilon_{i_{k-1}}, \varepsilon_{i_k}]$
and two neighboring bands $J_k$ and $J_{k+1}$ have a
common endpoint $\varepsilon_{i_k}$ if and only if there exist
$p_1$ and $p_2$ from $T^2$ such that $\widetilde{Q}(p_1;z)$ and
$\widetilde{Q}(p_2;z)$ do not have a pole at $\varepsilon_{i_k}$
and
  \begin{eqnarray*}
\widetilde{Q}(p_1;\varepsilon_{i_k})+\widetilde{A}(p_1) &\geq&
0\,,
\\ \widetilde{Q}(p_2;\varepsilon_{i_k})+\widetilde{A}(p_2) &\leq&
0\,.
  \end{eqnarray*}
There is at most one degenerate band corresponding to a constant
$E_{i_k}(\cdot)$. In particular, the degeneracy is excluded if the
matrix $A$ is diagonal, i.e. $A= \alpha I_{\ell^2(\Gamma)}$ for
some $\alpha\in\R$. The absolutely continuous spectrum of $H_A$ is
the union $\bigcup_{k=0}^\infty J_k$ with the exception of the
possible degenerate band.\\ [1mm]
(ii) The point part of the spectrum consists, in addition to the
mentioned degenerate band, of the modified Landau levels $z_0$
from $\sigma(H_0)$ which persist under the perturbation. This
concerns the whole $\sigma(H_0)$ if $N\ge 2$ while for $N=1$, the
levels $z_0 \in \sigma(H_0)$ for which there is just one pair of
indices $(l,n)\in J(z_0)$ and $\chi_n(\kappa_3) \neq 0$ have to be
removed.
  \end{theorem}
{\sl Proof:} Most part follows from Lemma~\ref{lemma: multiplicity
1} and Lemma~\ref{lemma: exten E 1}; it remains us to check the
claims about the degenerate band in (i). Suppose that
there are two different degenerate bands $\{E \}$ and $\{E' \}$
with $E,\,E'\in\R$ separated from the spectrum of $H_0$. Then we
have $\widetilde{Q}(p;E)+\widetilde{A}(p)=0$ for all $p \in T^2$
and the same for $E'$ which yields
  \begin{equation} \label{EE'}
Q(\vec \gamma, \vec \kappa;E)= - A(\vec \gamma,\vec \kappa)=
Q(\vec \gamma, \vec \kappa;E') \nonumber
  \end{equation}
for all $\vec \gamma$ from the set $\Lambda$. In particular,
choosing $\vec \gamma$ with a large modulus, we find different
terms for the corresponding matrix elements of $Q$ at different
energies $E$ and $E'$, because
  \begin{equation} \label{asyQ}
Q(\vec \gamma, \vec \kappa;E)= C(E) \, e^{-{|B| \over 4}
|\gamma|^2} |\gamma|^{ {E -|B|-({\pi \over d})^2 \over |B|}}
\left( 1 +\OO (|\gamma|^{-\nu}) \right) \nonumber,
  \end{equation}
where $\nu =\min \left\{ 2,{3 \over |B|}({\pi \over d})^2
\right\}$, which leads to a contradiction with (\ref{EE'}). The
expansion (\ref{asyQ}) follows from the definition of $Q$ by
(\ref{def Q}) in combination with the asymptotic behavior of the
hyper\-geometric function, $U(a,1;s)=s^{-a}(1+\OO(|s|^{-1}))$ --
see \cite[13.1.8]{AS}. Furthermore, consider a diagonal matrix $A$
and suppose that there is a degenerate band $\{ E\}$. Then the
condition $\widetilde{Q}(p;E)+\alpha=0$ holds for any $p \in T^2$.
In view of the relation (\ref{tilde Q}) it leads to the
requirement $Q(\vec \gamma, \vec \kappa;E)=0$ for all $\vec \gamma
\neq \vec \kappa$ which again contradicts the known asymptotic
behavior. \quad \QED \vspace{1em}

Thus we have obtained spectral bands between neighboring points of
the unperturbed spectrum as in the planar case \cite{G}. Needless
to say, the bands are not the same because the dispersion
functions are different and also the unperturbed spectrum is
different: its points are sums of the `two-dimensional' Landau
levels and energies of transverse modes. In this sense the band
structure in a layer is richer.

The most important difference from the planar case is the possible
existence of a spectral gap containing the whole interval
$(\varepsilon_{i-1},\varepsilon_i)$ for some integer $i$, so that
the free-spectrum gap is preserved by the perturbation. Such a
situation occurs, for example, if the positions of point
potentials coincide with a node of each transverse mode
corresponding to $\varepsilon_i$, i.e. $\chi_n(\kappa_3)=0$ for
all $(l,n) \in J(\varepsilon_i)$, and if at the same time
$\widetilde{Q}(p;\varepsilon_i) + \widetilde{A}(p) \leq 0$ holds
for all $p \in T^2$. The last condition is satisfied, e.g., for a
diagonal matrix $A=\alpha I$ with the parameter $\alpha \leq -
\max_{p \in T^2}|\widetilde{Q}(p;\varepsilon_i)|$ as it follows
from monotonicity of the function $\widetilde{Q}(p;\cdot)$.
Although it is not a generic situation it is not purely
hypothetical. On the other hand, such a preserved gap cannot occur
for the lowest interval, $k=0$, or in the situation when both
endpoints contain the contribution from the lowest transverse
mode, i.e. $(l,1)\in J(\varepsilon_{i-1})$ and $(l',1)\in
J(\varepsilon_i)$. In case of a thin layer, it means that there is
exactly one spectral band between each two neighboring modified
Landau levels below $\varepsilon(0,2)$, i.e. below the threshold
of the second transverse mode; it is obvious that the thinness
here has to be understood in comparison to the characteristic
length given by the magnetic field.


\setcounter{equation}{0}
\section{The case of integral flux and polyatomic crystal}

Consider next a polyatomic lattice $\Lambda$, i.e. suppose that
the set $K$ contains more than one point, $\vec \kappa_i \neq \vec
\kappa_j$ for $i,j=1,\ldots,|K|, \;i \neq j$. The flux $\eta$ is
again an integer number $N$. Compared to (\ref{Four}) the Fourier
transformation must be modified; ${\cal F}_\eta: \; \ell^2(\Gamma)
\mapsto L^2(T^2) \otimes \ell^2(K)$ acts now as
   \begin{equation}
({\cal F}_\eta \phi) (p;\vec \kappa) = \sum_{\vec \lambda \in
\Lambda} \phi(\vec \lambda) \exp \left[\pi i \xi \,
\kappa\wedge\lambda - 2\pi i (\lambda_a p_1 + \lambda_b p_2+ N
\lambda_a \lambda_b /2)\, \right],
   \end{equation}
i.e. values of the transformed function ${\cal F}_\eta \phi$ of
the variable $p\in T^2$ are no longer scalar but rather
$|K|$-dimensional vectors. The Fourier transformed operator ${\cal
F}_\eta D((\vec \lambda,1)) {\cal F}_\eta^{-1}$ with $\vec
\lambda\in\Lambda$ acts again as a multiplication by
$e_\lambda(p)$. An argument similar to that of the monoatomic case
leads to
   \begin{eqnarray}
\lefteqn{ (\widetilde{Q}(z)+\widetilde{A})(p;\vec \kappa, \vec
\kappa') = ({\cal F}_\eta (Q(z)+A) {\cal F}_\eta^{-1}) (p;\vec
\kappa, \vec \kappa) } \\ && \!\!\!\!\! = \sum_{\vec \lambda \in
\Lambda} (Q(z)+A)(\vec \lambda +\vec \kappa, \vec \kappa') \: \exp
\left[\pi i \xi \, \kappa\wedge\lambda - 2\pi i (\lambda_a p_1 +
\lambda_b p_2+ N \lambda_a \lambda_b /2) \right] \nonumber.
   \end{eqnarray}
The matrix elements $\widetilde{Q}(p; \vec \kappa, \vec
\kappa';z)$ are real-analytic with respect to $p\in T^2$ and
meromorphic in $z$ with simple poles which can be located only at
the points of $\sigma(H_0)$. The matrix elements $\widetilde{A}(p;
\vec \kappa, \vec \kappa')$ are real-analytic in $p$. Finally, the
transformed Green function reads
  \begin{eqnarray}\label{Green function n}
\lefteqn{ G(p;k,l,n;k',l',n';z) = \delta_{kk'} \delta_{ll'}
\delta_{nn'} {1 \over \varepsilon(l,n)-z} } \\ && \!\!\! - \,
\sum_{\vec \kappa, \vec \kappa' \in K}
 [\widetilde{Q}(p;z)+\widetilde{A}(p)]^{-1} (\vec \kappa,
\vec \kappa') \: {\widetilde{\delta}_\kappa (p;k,l) \over
\varepsilon (l,n)-z} \; {\widetilde{\delta}_{\kappa'}^* (p;k',l')
\over \varepsilon (l',n')-z} \, \chi_n(\kappa_3)
\chi_{n'}(\kappa_3') \nonumber\,.
  \end{eqnarray}
Using the known relation $d\widetilde{Q}/dz=
\widetilde{\Gamma}^*_{z^*} \widetilde{\Gamma}_z$ for the
derivative, we get
  \begin{equation}\label{derivative Q n}
{\partial \widetilde{Q} \over \partial z}(p; \vec \kappa, \vec
\kappa';z) = \sum_{l=0}^\infty \sum_{n=1}^\infty {1 \over
|\varepsilon(l,n)-z|^2} \; \chi_n(\kappa_3) \chi_n(\kappa_3') \,
\sum_{k=0}^{N-1} \widetilde{\delta}^*_\kappa (p;k,l)
\widetilde{\delta}_{\kappa'} (p;k,l) \nonumber.
  \end{equation}
The asymptotic behaviour of the diagonal elements of the matrix
$\widetilde{Q}(p;z)$ as $z \to -\infty$ is similar to (\ref{nonmagnetic asymp}),
while the non-diagonal elements are bounded,
  \begin{equation}\label{nonmagnetic asymp n}
\widetilde{Q}(p; \vec \kappa, \vec \kappa';z) = - \delta_{\vec
\kappa \vec \kappa'} \, {\sqrt{-z} \over 4\pi} +\OO(1)\,.
  \end{equation}

Let us begin the spectral analysis with the fiber operator
$\widetilde{H}_A(p)$ for a fixed $p \in T^2$. For each point
$\varepsilon_i \in \sigma(H_0)$ we define matrix
$G_{\varepsilon_i} (p)$ as the residue of $-\widetilde{Q}(p;z)$ at
$z=\varepsilon_i$, i.e.
  \begin{equation}
G_{\varepsilon_i}(p; \vec \kappa, \vec \kappa') = \sum_{(l,n) \in
J(\varepsilon_i)} \chi_n(\kappa_3) \chi_n(\kappa_3')
\sum_{k=0}^{N-1} \widetilde{\delta}_\kappa^* (p;k,l)
\widetilde{\delta}_{\kappa'} (p;k,l)\,,
  \end{equation}
and we denote its rank by $r_{\varepsilon_i}(p)$. We further
define $P_{\varepsilon_i}(p)$ as the orthogonal projection onto
$\ker G_{\varepsilon_i}(p) \subset \ell^2(K)$ and operator
$D_{\varepsilon_i}(p)$ as
  \begin{equation}
D_{\varepsilon_i}(p) = \lim_{z\to \varepsilon_i}
P_{\varepsilon_i}(p) (\widetilde{Q}(p;z)+\widetilde{A}(p)) |_{\ker
G_{\varepsilon_i}(p)}\,.
  \end{equation}

The multiplicity of $\varepsilon_i$ in the spectrum of the free
Hamiltonian is equal to $N |J(\varepsilon_i)|$ with the eigenspace
$\Omega(\varepsilon_i)$ defined by (\ref{eigenspace}). The second
term of the Green function (\ref{Green function n}) modifies the
residue at $z=\varepsilon_i$ in two possible ways: \\ [1mm]
(i) taking into account contribution from the indices $(l,n),\,
(l',n') \in J(\varepsilon_i)$ to the kernel of the resolvent, we
get the orthogonal projection to the subspace spanned by the
vectors $( \chi_{n_i}(\kappa_3) \widetilde{\delta}_\kappa
(p;k,l_{i}) )_{k,i}$ with $k=0,\,\ldots,N-1\,$ and $(l_i,n_i)\in
J(\varepsilon_i)$. In this way the multiplicity of $\varepsilon_i$
is diminished by $r_{\varepsilon_i}(p)$.\\ [1mm]
(ii) On the other hand, for indices $(l,n) \notin
J(\varepsilon_i)$ we get a nonzero residue when the operator
$D_{\varepsilon_i}(p)$ is not invertible. The corresponding
eigenspace is orthogonal to $\Omega(\varepsilon_i)$ with the
maximal possible dimension equal to $|K|$.
\vspace{1mm}

Putting two terms of the Green function (\ref{Green function n})
together, we arrive at the following result:
\begin{lemma}\label{lemma: multiplicity n}
Assume that the flux $\eta$ equals an integer number $N$ and the
elementary cell of the lattice $\Gamma$ contains $|K|$ point
interactions. Choose a point $\varepsilon_i\in \sigma(H_0)$ and
fix $p \in T^2$. Then the multiplicity $d_{\varepsilon_i}(p)$ of
$\varepsilon_i$ in $\sigma(\widetilde{H}_A(p))$ is equal to
$$ d_{\varepsilon_i}(p) = N |J(\varepsilon_i)| -
r_{\varepsilon_i}(p), $$
if $D_{\varepsilon_i}(p)$ is invertible, while in the opposite
case it satisfies the inequalities
$$ N |J(\varepsilon_i)| - r_{\varepsilon_i}(p) \leq
d_{\varepsilon_i}(p) \leq N |J(\varepsilon_i)| + |K|
-r_{\varepsilon_i}(p)\,. $$
\end{lemma}
\vspace{1mm}

Apart from the modified Landau levels of the free system, the
spectrum of $\widetilde{H}_A(p)$ contains eigenvalues due to the
presence of the point potentials. A necessary condition for $E
\in\R \setminus \sigma(H_0)$ to be an eigenvalue is
  \begin{equation}\label{implicit eq n}
\det [\widetilde{Q}(p;E)+ \widetilde{A}(p)] =0.
  \end{equation}
Notice that $E$ might not be an eigenvalue of $\widetilde{H}_A(p)$
if the vectors
$$ \psi_{\vec \kappa} = \Big( {\widetilde{\delta}_\kappa (p;k,l)
\over \varepsilon(l,n) - E} \, \chi_n(\kappa_3) \Big)_{k,l,n} \,,
\qquad \vec \kappa \in K$$
were not linearly independent, in which case the second term in
Green function (\ref{Green function n}) could vanish. This
cannot happen, however, because $\partial \widetilde{Q}/ \partial
z$ is the Gram matrix for this $|K|$-tuple of vectors and by
\cite{KL} one has for a fixed $z\not\in \sigma(H_0)$ the
inequality
  \begin{equation}\label{positivity}
{\partial \widetilde{Q}(p;z) \over \partial z} \geq c_z
I_{\ell^2(K)}
  \end{equation}
with some $c_z >0$. Therefore $\psi_{\vec \kappa}$ are linearly
independent and eigenvectors corresponding to $E$ are given by $
\sum_{\vec \kappa \in K} \beta_{\vec \kappa} \psi_{\vec \kappa}$,
where the vectors $\{\beta_{\vec \kappa}\}$ belong to $\ker
[\widetilde{Q}(p;E) + \widetilde{A}(p)]$.

A useful way to solve the implicit equation (\ref{implicit eq n})
is by examining the eigenvalues $\mu^{(j)}(p;z)$, $j=1,\ldots,|K|$
of the matrix $\widetilde{Q}(p;z) + \widetilde{A}(p)$ with the
numbering which takes their multiplicity into account. Apparently,
a solution of equation $\mu^{(j)}(p;E)=0$ for some $j$ solves also
the original equation. Properties of the functions $\mu$ are
described in following lemma:
  \begin{lemma}
Suppose that the flux $\eta$ is an integer number $N$ and the
elementary cell of $\Gamma$ contains $|K|$ point potentials, and
fix $p \in T^2$. Then the eigenvalues $\mu^{(j)}(p;z)$ of the
matrix $\widetilde{Q}(p;z) + \widetilde{A}(p)$ are monotonously
increasing functions of $z$ in each interval
$(\varepsilon_{i-1},\varepsilon_i)$, $i=0,1,\ldots\,$. When $z$
approaches a modified Landau level $\varepsilon_i$, exactly
$r_{\varepsilon_i}(p)$ functions of the family
$\mu^{(j)}(p;\cdot)$, $j=1,\ldots,|K|$ diverge, while the
remaining ones have finite limits. When $z$ tends to $-\infty$,
all the $|K|$ functions $\mu^{(j)}(p;\cdot)$ diverge to $-\infty$.
  \end{lemma}
{\sl Proof:} Since the matrix $\widetilde{Q}(p;z_2)
-\widetilde{Q}(p;z_1)$ with $\varepsilon_{i-1}<z_1<z_2<
\varepsilon_i$ is by (\ref{positivity}) positive definite, the
Lidskii's theorem \cite[Thm.~II.6.10]{K} yields the inequality
$\mu^{(j)}(p;z_1)< \mu^{(j)}(p;z_2)$ for each $j$, hence all the
functions $\mu^{(j)}(p;\cdot)$ increase monotonously between every
two neighboring points of $\sigma(H_0)$. To prove the second
claim, take $|K|$ eigenvalues $\beta^{(j)}(p;z)$,
$j=1,\ldots,|K|$, of matrix $(\varepsilon_i-z) [\widetilde{Q}(p;z)
+ \widetilde{A}(p)]$. By Rellich's theorem \cite[Thm.~II.6.8]{K}
the functions $\beta^{(j)}$ are continuously differentiable in
some neighborhood of $\varepsilon_i$, an for $z=\varepsilon_i$ the
matrix under considerations coincides with $G_{\varepsilon_i}$.
Finally, using Gershgorin's circles -- see \cite[Thm.~XIV.5.5]{Ga}
-- in combination with the asymptotic formula (\ref{nonmagnetic
asymp n}) for the matrix $\widetilde{Q}(p;z)$ we get the last
claim. $\;$ \QED \vspace{1em}

The lemma implies that there are at most $|K|$ eigenvalues in each
interval $(\varepsilon_{i-1},\varepsilon_i)$, $i=0,1,\ldots$ which
is after all clear from general principles --
cf.~\cite[Sec.~8.3]{We}. To get a more specific information
about the actual number of eigenvalues, we have to look closely at
the number $r_{\varepsilon_i}(p)$.
  \begin{remark}\label{remark: gram matrix}
{\rm Since the matrix $G_{\varepsilon_i}(p)$ is obviously the Gram
matrix of the system of $N |J(\varepsilon_i)|$-dimensional vectors
$$ \Big( \chi_{n_s}(\kappa_3) \widetilde{\delta}_\kappa
(p;k,l_{s}) \Big)_{k,s} \qquad  k=0,\ldots,N-1\,, \; (l_s,n_s)\in
J(\varepsilon_i)\,, $$
the rank $r_{\varepsilon_i}(p)$ cannot exceed $\min(|K|,N
|J(\varepsilon_i)|)$. There are three situations when the maximum
value specified here cannot be achieved whatever point $p$ is
considered. First, the dimension of the vectors is in fact smaller
if $\chi_{n_s}(\kappa_3)=0$ holds for some index $n_s$ and all
$\vec \kappa \in K$. In general, the said dimension is equal to
$N|\tilde{J}(\varepsilon_i)|$, where $\tilde{J}(\varepsilon_i)$ is the set
$J(\varepsilon_i)$ from which the pairs $(l_s,n_s)$ with the
described property were deleted. Second, the number of the vectors
is $|K|$, but when a lattice point $\vec \kappa$ is such that
$\chi_{n_s}(\kappa_3) =0$ holds for all $(l_s,n_s) \in
J(\varepsilon_i)$, the corresponding vector is zero for any $p$.
Excluding such elements from the set $K$ we obtain a subset
denoted as $\bar K(\varepsilon_i)$.

Finally, we must also examine carefully the situation when several
points $\vec \kappa^j$, $j=1,\ldots,q$, from the elementary cell
differ in the third coordinate only, i.e. they are arranged
vertically in the layer. If the level $\varepsilon_i$ is not
degenerate, by neglecting all but one of the $q$ vectors in
question, we do not change the rank of the Gram matrix, provided
the one which we keep is not a zero vector. If a degenerate level
is admitted, then the number of linearly independent vectors among
these $q$ vectors is less or equal to the rank of the following
matrix
$$ \Big( \chi_{n_s}(\kappa^j_3) \Big)_{j,s} \qquad j=0,\ldots,q,
\; (l_s,n_s)\in \tilde{J}(\varepsilon_i)\,. $$
Then we eliminate the remaining ones of the $q$ vectors from
$\bar K(\varepsilon_i)$ and obtain in this way another subset of
$K$ which we denote as $\tilde K(\varepsilon_i)$. The maximal possible
rank $r_{\varepsilon_i}(p)$ of the matrix $G_{\varepsilon_i}(p)$
is in this way equal to
$$\min(|\tilde K(\varepsilon_i)|,N |\tilde J(\varepsilon_i)|)\,;$$
whether the maximum is achieved at a point $p$ or not now depends
on the functions $\widetilde{\delta}_\kappa (\cdot;k,l)$. }
  \end{remark}

Such a dependence of $r_{\varepsilon_i}(p)$ on the parameters of
the problem makes the general spectral analysis cumbersome. In
what follows we will thus restrict our attention to the generic
situation only and we impose additional restrictions on the
Hamiltonian $H_A$. First of all, we assume that $(\pi /d)^2$ and
$|B|$ are not rationally related, so that $J(\varepsilon_i)=1$
holds for all modified Landau levels $\varepsilon_i$. Furthermore,
we define the sets
\begin{eqnarray*}
U'_{\varepsilon_i} &\!:=\!& \{ p \in T^2: \;
r_{\varepsilon_i}(p)=r_{\rm max}\equiv \min(N,|\tilde K|) \}\,, \\
U''_{\varepsilon_i} &\!:=\!& \{ p \in T^2: \; D_{\varepsilon_i} \;
{\rm \mathrm{is \; invertible}} \}
\end{eqnarray*}
and
  \begin{equation}\label{set U n}
U_{\varepsilon_i}:=U'_{\varepsilon_i} \cap U''_{\varepsilon_i}\,,
  \end{equation}
where $|\tilde K|$ represents the number of points $\vec \kappa \in K$
after we have excluded $q-1$ points from every $q$-tuple in which
the first two coordinates coincide. The set $\tilde K$ does not depend
on $\varepsilon_i$ any more.

In the rest of the paper, we consider only operators $H_A$ such
that the set $\bigcap_{i\geq 0} U_{\varepsilon_i}$ is nonempty,
which is a generic situation. One can check easily
that by an arbitrarily small shift of the points $\vec \kappa$ in
the elementary cell $Q_{\Lambda}$ we can achieve that
$\bigcap_{i\geq 0} U'_{\varepsilon_i} \neq \emptyset$, and in a
similar way, by an arbitrarily small perturbation of the diagonal
elements of the matrix $A$ we can always satisfy the condition
$\bigcap_{i\geq 0} U''_{\varepsilon_i} \neq \emptyset$. Recall
that the possible positions $\vec \kappa$ of point potentials in
the elementary cell are dense in $Q_\Lambda^{|K|}$, while all
possible values of $A(\vec \kappa, \vec \kappa)$ span $\R^{|K|}$,
and consequently, the set
$$ \left\{ (\vec \kappa_1,\ldots,\vec \kappa_{|K|},A(\vec
\kappa_1,\vec \kappa_1), \dots,A(\vec \kappa_{|K|},\vec
\kappa_{|K|})) \in Q_\Lambda^{|K|} \times \R^{|K|} \, : \;
\bigcap_{i\geq 0} U_{\varepsilon_i} \neq \emptyset  \right\} $$
has a full measure in $Q_\Lambda^{|K|} \times \R^{|K|}$.

For $|K|=r_{\rm max}$ and $p \in U_{\varepsilon_{i-1}} \cap
U_{\varepsilon_i}$, the number of the eigenvalues in the free
Hamiltonian spectral gap equals $|K|$. Now we employ the
assumption $\bigcap_{i\geq 0} U_{\varepsilon_i} \neq \emptyset$:
since $\widetilde{\delta}_\kappa (p;k,l)$ and $\widetilde{Q}(p;z)$
with $z \notin \sigma(H_0)$ are analytic as functions of $p$, all
sets the $U'_{\varepsilon_i}$ and $U''_{\varepsilon_i}$, $\:i \geq
0$, have full measures, and the same is true for the intersection
$U_{\varepsilon_{i-1}} \cap U_{\varepsilon_i}$. The functions
$E_i^{(j)}(\cdot)$ are bounded, which follows for $i=0$ from the
asymptotic formula (\ref{nonmagnetic asymp n}) and the boundedness
of the matrix $\widetilde{A}$ while otherwise the claim is valid
trivially. Thus we may extend the $|K|$ dispersion functions
$E_i^{(j)}(\cdot)$ by continuity to the entire torus $T^2$. We
arrive at the following result.
\begin{lemma}
Assume that $|K|=r_{\rm max}$ and the Hamiltonian $H_A$ satisfies
$\bigcap_{i\geq 0} U_{\varepsilon_i} \neq \emptyset$. Consider the
extended functions $E_i^{(j)}(p)$, $j=1,\ldots,|K|$, defined on
$T^2$ in the described way. Then the inequalities
$\varepsilon_{i-1}<E_i^{(j)}(p)<\varepsilon_i$ for some $j$, $1
\leq j \leq |K|$, implies that $E_i^{(j)}(p)$ is a solution of the
implicit equation (\ref{implicit eq n}).
\end{lemma}

For $|K|>r_{\rm max}$ and $p \in U_{\varepsilon_{i-1}} \cap
U_{\varepsilon_i}$, the number of eigenvalues in the interval
$(\varepsilon_{i-1},\varepsilon_i)$ is not necessarily equal to
$|K|$. Consider first $p \in \bigcap_{i\geq 0} U_{\varepsilon_i}$.
In this case, $|K|-r_{\rm max}$ functions $\mu^{(j)}(p;z)$ do not
diverge when $z$ approaches an endpoint of the interval, instead
they `meet' there one of functions $\mu$ from the neighboring
interval. Then it is natural to unify these functions coming from
neighboring intervals obtaining new functions $\nu_i^{(j)}(p;)$,
$i=0,1,\ldots\,$, $j=1,\ldots,r_{\rm max}$, which are defined on
the enlarged intervals $(\varepsilon_{i-s},\varepsilon_i)$, where
$1 \leq s \leq |K|/r_{\max}$ if the fraction is integer, otherwise
$1 \leq s \leq [|K|/r_{\rm max}]+1$. We put
$\varepsilon_{i-s}=-\infty$ and $U_{\varepsilon_{i-s}}=T^2$
whenever $i-s<0$. On each subinterval, the corresponding families
of functions $\nu(p;z)$ and $\mu(p;z)$ coincide. The function
$\nu_i^{(j)}(p;z)$ tends to $\pm \infty$ as $z$ approaches
$\varepsilon_i$ or $\varepsilon_{i-s}$, respectively. Thus there
exists exactly one solution to the equation $\nu_i^{(j)}(p;E)=0$
which we denote as $E_i^{(j)}(p)$. The same conclusion can be made
under a weaker assumption, $p \in \bigcap_{l=0}^{s}
U_{\varepsilon_{i-l}}$. This set has again a full measure and
functions $E_i^{(j)}(\cdot)$ are bounded, so we can extend the
dispersion functions by continuity to the entire torus $T^2$.

After having analyzed the fiber operator spectrum, let us run $p$
run through the torus $T^2$ to get the spectrum of the Hamiltonian
$H_A$.
\begin{theorem}\label{thm: spectrum n}
Suppose that the flux $\eta$ is an integer number $N$ and the
elementary cell of the lattice $\Gamma$ contains $|K|$ points. In
addition, let $(\pi / d)^2$ and $|B|$ be irrationally related and
let the Hamiltonian $H_A$ satisfy the condition $\bigcap_{i\geq 0}
U_{\varepsilon_i} \neq \emptyset$, where the sets
$U_{\varepsilon_i}$ are defined by (\ref{set U n}). Then the
spectrum of $H_A$ consists of two parts, namely \\ [1mm]
(i) spectral bands $I_i^j$ with $\,j=1,\ldots,r_{\rm max}$ and
$i=0,1,\ldots\,$, where $r_{\rm max}=\min(N,|\tilde K|)$ and
$\tilde K$ is the
maximal subset of $K$ such that no pair of points from $K'$
coincide in the first two coordinates. Each band $I_i^j$ is given
as the range of the extended function $E_i^{(j)}(p)$, $p \in T^2$,
which is defined by the implicit equation (\ref{implicit eq n});
it lies within the interval $[\varepsilon_{i-s}, \varepsilon_i]$,
where $1 \leq s \leq |K|/r_{\max}$ if the fraction is integer and
$1 \leq s \leq [|K|/ r_{\rm max}]+1$ otherwise. The absolutely
continuous spectrum of $H_A$ is the union $\bigcup_{i=0}^\infty
\bigcup_{j=1}^{r_{\rm max}} I_i^j$ except possible bands
degenerate to a point, \\ [1mm]
(ii) modified Landau levels from the spectrum of $H_0$. If $N \leq
r_{\rm max}$, some points of $\sigma(H_0)$  may be absent from the
spectrum of $H_A$.
\end{theorem}

The theorem says nothing about possible common endpoint of two
neighboring bands, which coul be compared to the analogous part of
Theorem~\ref{thm: spectrum 1} in the monoatomic case. However, the
term `neighboring bands' does not have much sense here; the bands
$I_i^j$ with the same index $i$ may overlap and for $|K|>r_{\rm
max}$ also bands with different indices $i$ may overlap.

Apart from the difference between the modified Landau levels here
from the unperturbed spectrum in the planar case, the structure of
the two spectra is similar with one difference: the number of
bands $I_i^j$ which neighbor with the same point $\varepsilon_i$
from above equals $\min(N,|\tilde K|)$, while in the planar case
\cite{G} it is $\min(N,|K|)$ instead. The reason is clear: the
magnetic field perpendicular to the layer does not `distinguish'
two points placed one on the top of the other, and such a
situation can never occur in the planar case.

In the previous chapter we have found that the spectrum can
contain a gap covering the whole interval $(\varepsilon_{i-1},
\varepsilon_i)$ for some $i$. In the polyatomic case it is
obviously possible for $|K| \geq 2r_{\rm max}$, while the above
discussion excludes such a situation otherwise. It might thus seem
that for $|K|=r_{\rm max}=1$ we get a contradiction. However, the
discrepancy comes from the stronger restriction we have imposed
upon the operator $H_A$; the condition $\bigcap_{i\geq 0}
U_{\varepsilon_i} \neq \emptyset$ does not allow
$\chi_n(\kappa_3)=0$ to hold for all $(l,n) \in J(\varepsilon)$.


\setcounter{equation}{0}
\section{The case of a rational flux}

The general case when the flux is a rational number, $\eta={N
\over M}$, can be in some sense reduced to the previous analysis.
We can pass to an integral flux $\eta'=N$ by enlarging the
elementary cell. The new lattice $\Lambda'$ is generated by
vectors $\vec a$ and $M \vec b$ and the new set $K'$ is given by
$K+\{0, \vec b,\ldots, (M-1)\vec b \}$.

However, the result obtained by this simple trick is not fully
correct because it does not take the $M$-fold degeneracy into
account. Recall that the Hamiltonian $H_A$ commutes with all
magnetic translations from $W_\eta$. Hence any of its eigenspaces
can be written as a direct sum of spaces of the irreducible
representations of $W_\eta$, which are $M$-dimensional. Therefore
additional modifications are needed here. We define the Fourier
transformation ${\cal F}_\eta: \; \ell^2(\Gamma) \mapsto
L^2(T^2_\eta) \otimes \C^M \otimes \C^M \otimes \ell^2(K)$ by the
prescription
   \begin{eqnarray}
({\cal F}_\eta \phi) (p;j,m,\vec \kappa) &=& \sum_{\lambda_a,
\lambda_b \in \Z} \phi(\lambda_a \vec a + (\lambda_b M +m) \vec b
+\vec \kappa)
\\ && \times \exp \Big[\pi i \xi \, \kappa\wedge(\lambda_a \vec a +
\lambda_b M \vec b) \Big] \nonumber \\ && \times \exp \Big[ - 2\pi
i \Big(\lambda_a p_1 + \lambda_b p_2+ {N \over 2} \,  \lambda_a
\Big(\lambda_b + {m+2j \over M} \Big) \Big) \Big]\,. \nonumber
   \end{eqnarray}
The transformed quantity ${\cal F}_\eta D(\cdot) {\cal
F}_\eta^{-1}$ is then a direct integral of multiples of
irreducible representations,
$$ {\cal F}\eta D(\cdot) {\cal F}_\eta^{-1} =
\int_{T^2_\eta}^\oplus d^2p\: \triangle' (\cdot;p) \otimes
I_{\ell^2(K)}\,. $$
The representation $\triangle_d (\cdot;p)$ acting on $\C^M \otimes
\C^M$ is given by
   \begin{eqnarray}
\triangle_d ((\vec a,1);p) &=& \triangle ((\vec a,1);p) \otimes
I_{\C^M}, \nonumber \\
\triangle_d ((\vec b,1);p) &=& S \otimes \triangle'(p), \nonumber
   \end{eqnarray}
where $S$ and $\triangle'(p)$ are operators on $\C^M$ which act at
the basis vectors $e_j$, $j=0,\ldots,M\!-\!1\,,\,$ in following
way,
$$ S e_j=e_{j \ominus 1}\,, \qquad \triangle'(p) e_j = \exp(-2\pi
i p_2 \delta_{j,M-1}) e_{j \oplus 1}\,, $$
with $\oplus$ and $\ominus$ representing the sum and difference in
$\{0,\ldots,M-1\}$ modulo M. This representation is equivalent to
the representation $\triangle(\cdot;p) \otimes I_{\C^M}$.

Applying now the Fourier transformation to the operator $Q(z)+A$
we get
   \begin{eqnarray}
&& (\widetilde{Q}(z)+\widetilde{A})(p;j,m,\vec \kappa,j',m', \vec
\kappa') = \delta_{jj'}\sum_{\lambda_a,\lambda_b \in \Z} \exp
\left[\pi i \xi \, \kappa\wedge (\lambda_a \vec a + \lambda_b M
\vec b ) \right] \nonumber \\ && \phantom{AAAA} \times \exp \left[
- 2\pi i \left(\lambda_a p_1 + \lambda_b p_2+ {N \over 2} \,
\lambda_a \Big( \lambda_b + {m+2j \over M} \Big) \right) \right]
\nonumber
\\ && \phantom{AAAA} \times (Q(z)+A)(\lambda_a \vec a +
(\lambda_b M +m) \vec b + \vec \kappa, m' \vec b + \vec
\kappa')\,.
   \end{eqnarray}
Finally, the transformed Green function reads
  \begin{eqnarray}
&& G(p;j,k,l,n;j',k',l',n';z) \;=\; \delta_{jj'} \delta_{kk'}
\delta_{ll'} \delta_{nn'} {1 \over \varepsilon(l,n)-z} \\ &&
\phantom{AAAAAAAA}- \, \delta_{jj'} \sum_{\vec \kappa, \vec
\kappa' \in K} \sum_{m,m'=0}^{M-1}
 [\widetilde{Q}(p,j;z)+\widetilde{A}(p,j)]^{-1} (m,\vec \kappa,
m',\vec \kappa') \nonumber \\ && \phantom{AAAAAAAAAAAA} \times
{\widetilde{\delta}_{m,\kappa} (p;j,k,l) \over \varepsilon
(l,n)-z} \; {\widetilde{\delta}_{m',\kappa'}^* (p;j,k',l') \over
\varepsilon (l',n')-z} \, \chi_n(\kappa_3) \chi_{n'}(\kappa_3')
\nonumber\,,
  \end{eqnarray}
where
  \begin{equation}
\widetilde{\delta}_{m,\kappa} (p;j,k,l) = N^{-1/2}
\sum_{r=-\infty}^\infty \exp \left(2\pi i r {p_2+k \over N}\right)
\psi_0^* (m b + \kappa; p_1+\eta j+r,l)\,.
  \end{equation}
We have employed a relation similar to (\ref{shifted delta}) with
$\lambda \in \Lambda'$ and an additional parameter $j$, and the
$M$-dimensional representation $\triangle(\cdot,p)$ instead of the
character $\chi(\cdot;p)$.

Next, we proceed to the spectral analysis in the same way as we
did in the previous section. The implicit equation defining the
dispersion function $E_i^{(r,j)}(p)$ has the form
  \begin{equation}\label{implicit eq N/M}
\det [\widetilde{Q}(p;j;E) +\widetilde{A}(p;j)] = 0
  \end{equation}
with the dimension of the matrix being equal to $|K|M$. Due to the
equivalence of the two representations which we have mentioned
above, the matrices $\widetilde{Q}(p;j;z) +\widetilde{A}(p;j)$ for
different $j$ are unitarily equivalent. This is the source of the
$M$-fold degeneracy of the eigenvalues of $\widetilde{H}_A(p)$,
and thus also of the spectral bands of $H_A$.

To simplify the description of the spectrum $\sigma(H_A)$, we
consider in the following theorem a lattice with $|K|=1$.
\begin{theorem}
Assume that the flux $\eta$ is a rational number $N/M$ and the
elementary cell contains one point potential. In addition, let
$(\pi/d)^2$ and $|B|$ be irrationally related, and let the
Hamiltonian $H_A$ satisfy the condition $\bigcap_{i \geq 0}
\bigcap_{j=0}^{M-1} U_{\varepsilon_i}^j \neq \emptyset$, where the
sets $U_{\varepsilon_i}^j$ are defined in analogy with the
expression (\ref{set U n}). Then the spectrum $\sigma(H_A)$
consists of two parts:
\newline
(i) The first one is the union of spectral bands $I^r_i$ with
$r=1,\ldots,r_{\max}$ and $i=0,1,\ldots$, where $r_{\max}
=\min(M,N)$. The band $I^r_i$ is given as a range of the function
$p \mapsto E^{(r,j)}_i(p)$ with $p\in T^2_\eta$, defined by the
implicit equation (\ref{implicit eq N/M}) for some $j$. Each band
$I^r_i$ is $M$-times degenerated and it lies within the interval
$[\varepsilon_{i-s}, \varepsilon_i]$, where $1 \leq s \leq M$ if
$N=1$ and $1 \leq s \leq [M/N]+1$ otherwise. The absolutely
continuous spectrum is the union of all bands $I^r_i$ except
possible bands degenerated to a point.
\newline
(ii) The second part of the spectrum contains modified Landau
levels from $\sigma(H_0)$. In the case $N \leq M$ some points of
$\sigma(H_0)$ may be absent from the spectrum $\sigma(H_A)$.
\end{theorem}


\section{Survey of the results}

We have analyzed the spectrum of a Dirichlet layer with a periodic
array of point perturbations in presence of homogeneous magnetic
field. The generic picture we have obtained for a rational flux,
$\eta=N/M$, has some well-known features analogous to \cite{G}: in
the case of a single potential in the elementary cell, there are
$\min(M,N)$ spectral bands which split off each modified Landau
level $\varepsilon_i$, each band is $M$-times degenerated, and its
location is not necessarily restricted to the gaps adjacent to the
Landau level in question; more precisely, if $M>N$ it may spread
below $\varepsilon_{i-1}$. In the case of $n>1$ perturbations in
the elementary cell, the number of the spectral bands changes to
$\min(nM,N)$, while the $M$-fold degeneracy remains the same.

Apart of the magnetic field and the lattice spacing, the system
has another parameter, namely the layer width $d$. Its first and
most visible effect on the spectrum is that the Landau levels in
the unperturbed spectrum are combined with the energies of
transverse modes (that is what we mean by modified Landau levels),
and thus they are described by a pair of quantum numbers. Due to
this fact the results are similar to those for the two-dimensional
system of \cite{G} only in the generic situation when
$(\pi/d)^2/|B|$ is an irrational number and no point potential is
placed at a node of a transverse mode.

If these additional conditions are not satisfied, one has to
examine each Landau level separately as we did in
Remark~\ref{remark: gram matrix}. For example, if $(\pi/d)^2$ and
$|B|$ are rationally related, some Landau levels have an extra
degeneracy and the number of bands may increase, because one must
consider appropriate multiple of the integer $N$. On the other
hand, if we allow the sites of point potentials coincide with a
node of a transverse mode, the number of bands may decrease
because unperturbed levels do not ``feel'' the interaction.


\subsection*{Acknowledgement}

The authors thank V.A.~Geyler for useful comments. The research
was partially supported by GAAS grant A1048101.


\bibliographystyle{plain}

\end{document}